\documentclass[twocolumn]{jpsj2}

\usepackage{amsmath,amssymb}
\usepackage{bm}
\usepackage{graphicx,color}

\title{Conserving Gapless Mean-Field Theory of a Multi-Component Bose-Einstein Condensate}

\author{Yoshiyuki {\sc Kondo} and Takafumi {\sc Kita}}
\inst{Department of Physics, Hokkaido University, Sapporo 060-0810}

\recdate{\today}

\abst{%
We develop a mean-field theory for Bose-Einstein condensation of spin-1 atoms with 
internal degrees of freedom.
It is applicable to nonuniform systems at finite temperatures with
a plausible feature of satisfying the Hugenholtz-Pines theorem and 
various conservation laws simultaneously.
Using it, we clarify thermodynamic properties and
the excitation spectra of a uniform gas.
The condensate is confirmed to remain in the same internal state from $T\!=\!0$ 
up to $T_{c}$ 
for both antiferromagnetic and ferromagnetic interactions.
The excitation spectra of the antiferromagnetic (ferromagnetic) interaction
are found to have only a single gapless mode,
contrary to the prediction of the Bogoliubov theory where three (two) of them
are gapless.
We present a detailed discussion on 
those single-particle excitations in connection 
with the collective excitations.
}

\kword{%
Bose-Einstein condensation, mean-field theory, spinor BEC,
Luttinger-Ward functional, Bogoliubov-de Gennes equation}

\begin{document}
\maketitle
\section{Introduction}
In 1998, Stamper-Kurn {\it et al.} \cite{Stamper-Kurn} have 
successfully confined 
$^{23}$Na atoms in an optical dipole trap. 
They thereby realized Bose-Einstein condensation (BEC)
with internal degrees of freedom corresponding to 
 the three hyperfine spin states $|F\!=\!1,m_F\!=\!\pm 1,0 \rangle$.
Up to now, multi-component BEC has been observed also in 
other systems such as $^{23}$Na ($F\!=\!2$)\cite{Gorlitz} and 
$^{87}$Rb ($F\!=\! 1$\cite{Baarrett,Erhard} and 
$F\!=\!2$\cite{Chang, Schmal}).
The internal degrees of freedom are expected to bring new physics
into BEC absent in magnetically trapped single-component systems.

Several theories have been constructed for the weakly interacting Bose gases
of the single-component system.
The basic Bogoliubov theory\cite{Bogoliubov,Fetter} 
is a perturbation theory without self-consistency 
which is applicable only near $T\!=\!0$.
The Gross-Pitaevskii (GP) equation\cite{Gross1,Gross2,Pitaevskii}
for the condensate may be regarded as
the inhomogeneous Bogoliubov theory without the quasiparticle contribution.
Among extensions of the Bogoliubov theory
to finite temperatures is the mean-field
Hartree-Fock-Bogoliubov (HFB) theory
derived with the Wick decomposition procedure.
However, it predicts unphysical energy gap in the excitation spectrum
in contradiction to the Hugenholtz-Pines theorem.\cite{Hugenholtz}
To remove it, one generally introduces another approximation called
the Shohno (``Popov'') approximation\cite{Shohno64,Kita-FP} of
neglecting the anomalous pair correlation completely.
However, the approximation 
has a fundamental flaw of yielding dynamical equations of motion 
which does not satisfy various conservation laws.
Thus, there are still no systematic self-consistent approximation
schemes in condensed Bose systems which carry the two properties
of the exact theory simultaneously:\cite{Hohenberg}
``gapless'' and ``conserving.''

Theoretical studies of multi-component BEC was started
by Ohmi and Machida\cite{Ohmi-Machida} and Ho\cite{Ho}
independently in 1998. 
They were based on the GP equation
for the spin-1 Bose gas to determine the structure of the
condensate wave function at $T\!=\!0$.
They have also clarified the excitation spectra by considering fluctuations
around the solution of the GP equation.
These pioneering works have been followed by detailed studies on
the equilibrium and dynamical properties of spin-$1$ 
BEC.\cite{Law,Pu,Wenxian1,cond-mat/0507521,Huang,Isoshima,Wenxian2}
However, most of them consider only the region near $T\!=\!0$
based on the Bogoliubov or GP equations.
In addition, other few studies at finite temperatures 
adopt the Shohno approximation which may not provide reliable
predictions with the reasons mentioned above,
especially for multi-component systems.
Thus, spin-$1$ BEC at finite temperatures is 
theoretically not well understood 
yet.

Recently, one of the authors has constructed a mean-field theory
for a single-component BEC at finite temperatures\cite{Kita-LT,Kita-FP}
which is both ``gapless'' and  ``conserving'' with 
finite anomalous pair correlation.
We here extend the theory to the three-component BEC.
We then apply it to a uniform gas with a contact interaction
to reveal its basic features as a function of temperature.
As mentioned above, 
no detailed studies have been made at finite temperatures even for 
the uniform system.
Especially, we wish to clarify the condensate wave function and
the excitation spectra of the
multi-component system at finite temperatures
which may have non-trivial structures.

This paper is organized as follows. 
Section 2 derives a closed set of equations to determine the thermodynamic equilibrium
of general nonuniform three-component BEC.
Section 3 applies the formulation to a uniform gas under constant density.
Section 4 presents numerical results for  
spin-dependent antiferro- and ferromagnetic interaction.
Section 5 concludes the paper.
In Appendix, we present the multi-component
version of the Hugenholtz-Pines theorem.\cite{Hugenholtz}
We put $\hbar \!=\! k_{\rm B} = 1$ throughout.

\section{Formulation} 
\subsection{Hamiltonian}
We consider a system described by the Hamiltonian:
\cite{Ho,Ohmi-Machida}
\begin{align}
        &\mathcal H = \sum_{\sigma}
        \int {\rm d} \bm r \,\psi^\dagger_\sigma(\bm r) (H_0 - \mu) \psi_\sigma(\bm r)
         + \frac{1}{2}\sum_{\sigma\sigma'\tau\tau'}
         \int {\rm d}\bm r_1 {\rm d} \bm r_2 \notag\\
		  &\hspace{7mm}\times 
		  \mathcal{U}_{\sigma\sigma',\tau\tau'}(\bm r_1 \!-\! \bm r_2) 
         \psi^\dagger_{\sigma}(\bm r_1) \psi^\dagger_{\tau}(\bm r_2) 
         \psi_{\tau'}(\bm r_2) \psi_{\sigma'}(\bm r_1) \, .
        \label{eq:hamiltonian}
\end{align}
Here $H_0 \!\equiv\! - \frac{\nabla^2}{2M} \!+\! V_{\rm ext}(\bm r)$
with $M$ the mass and $V_{\rm ext}$ the external potential, 
$\mu$ is the chemical potential, and
$\psi_{\sigma}(\bm r)$ is the field operator
with the subscript $\sigma\!=\!0,\pm 1$ denoting the spin component. 
The quantity $\mathcal{U}_{\sigma\sigma',\tau\tau'}$ is a spin-dependent interaction
given by
\begin{align}
        &\mathcal{U}_{\sigma\sigma',\tau\tau'}(\bm r_1 - \bm r_2) 
        \notag \\
        &= \delta(\bm r_1 - \bm r_2)
        \left( c_0 \delta_{\sigma\sigma'} \delta_{\tau\tau'} 
        + c_2 \bm F_{\sigma\sigma'} \cdot \bm F_{\tau\tau'} \right),
		  \label{eq:U_ijmn}
\end{align}
where $c_0$ and $c_2$ are parameters
defined in terms of spin dependent $s$-wave scattering length $a_f$ 
($f\!=\!0,1,\cdots$) as\cite{Ho}
\begin{subequations}
	\begin{align}
		  c_0 &= \frac{4 \pi}{M}\frac{(a_0+2a_2)}{3} \, ,
		  \\
		  c_2 &= \frac{4 \pi}{M}\frac{(a_2 - a_0)}{3} \, ,
	\end{align}
		  \label{c_0c_2}
\end{subequations}
and $\underline{\bm F}\!=\!({\bm F}_{\sigma\sigma'})$ is the spin matrix for $F=1$: 
\begin{subequations}
\begin{align}
        \underline{F}^1 &=\frac{1}{\sqrt{2}}
              \begin{bmatrix} 0 & 1 & 0 \\
                              1 & 0 & 1 \\
                              0 & 1 & 0 
              \end{bmatrix}
              ,\\
        \underline{F}^2 &= \frac{i}{\sqrt{2}}
              \begin{bmatrix} 0 & -1 & 0 \\
                              1 & 0 & -1 \\
                              0 & 1 & 0 
              \end{bmatrix}
              ,\\
        \underline{F}^3 &= 
        \begin{bmatrix}
                1 & 0 & 0 \\
                0 & 0 & 0 \\
                0 & 0 & -1
        \end{bmatrix} .
        \label{}
\end{align}
\label{underlineF}
\end{subequations}

\noindent
The second term in the round bracket of eq.\ (\ref{eq:U_ijmn}) is similar in form with 
the Heisenberg spin Hamiltonian. 
We hence call the $c_2\!>\!0$ ($c_2\!<\!0$) case as antiferromagnetic
(ferromagnetic).

It is convenient to introduce the Nambu vector:
\begin{align}
 \hat{\bm\psi}^{\dagger} = 
        [\psi_1^\dagger \ \psi_0^\dagger \ \psi_{-1}^\dagger \ \psi_1 
        \ \psi_0 \ \psi_{-1}]\, , \hspace{5mm}
        \hat{\bm\psi} = 
        \begin{bmatrix}
        \psi_1 \\ \psi_0 \\ \psi_{-1} \\ \psi_1^\dagger 
        \\ \psi_0^\dagger \\ \psi_{-1}^\dagger
        \end{bmatrix} .
        \label{eq:Nambu}
\end{align}
Using eq.\ (\ref{eq:Nambu}) and 
the normal-ordering operator $\mathcal{N}$,\cite{Fetter-Walecka}
the interaction in eq.\ (\ref{eq:hamiltonian})
is transformed into
\begin{align}
		  \mathcal{H}'= \frac{1}{8} \sum_{\nu=0 }^{3}
		  & \int {\rm d} \bm r {\rm d}\bm r' \mathcal{U}_{\nu}(\bm r - \bm r') 
		  \notag\\
		  & \times \mathcal{N}
		  [\hat{\bm\psi}^\dagger(\bm r)   \hat{A}^{\nu} \hat{\bm\psi}(\bm r)
		   \hat{\bm\psi}^\dagger(\bm r')  \hat{A}^{\nu} \hat{\bm\psi}(\bm r')] \, .
        \label{H'}
\end{align}
Here $\mathcal{U}_{\nu}$ is defined by
\begin{align}
\mathcal{U}_{\nu}(\bm r - \bm r')\!=\! \left\{
\begin{array}{ll} c_0 \delta(\bm r - \bm r') & : \nu=0 \\
c_2 \delta(\bm r - \bm r') & : \nu=1,2,3
\end{array}
\right.
\, ,
\label{U_nu}
\end{align}
and $\hat{A}^{\nu}$ denotes the vertex:
\begin{align}
\hat{A}^{\nu}=\begin{bmatrix} \underline{A}^{\nu} & \underline 0 \\  \underline 0 & 
(\underline{A}^{\nu})^{\rm T} 
\end{bmatrix}\, ,
\end{align}
with $\underline{A}^{\nu}$ defined 
in terms of the unit matrix $\underline{1}$ and eq.\ (\ref{underlineF}) as
\begin{align}
\underline{A}^{\nu}\!=\! \left\{
\begin{array}{ll} \underline{1} & : \nu=0 \\
\underline{F}^{\nu} & : \nu=1,2,3
\end{array}
\right.
\, .
\label{A_nu}
\end{align}
The equivalence between the interaction term in eq.\ (\ref{eq:hamiltonian}) and 
eq.\ (\ref{H'}) can be checked
easily by writing the latter without using $\mathcal{N}$.
The expression (\ref{H'}) has an advantage that the perturbation expansion
with respect to $\mathcal{H}'$ can be carried out directly by using the 
Nambu Green's function; see Appendix A of ref.\ \citen{Kita-FP}
for details.

\subsection{Mean-field equations}

To describe the condensed phase, let us express 
the field operator ${\bm\psi}(\bm r)\!\equiv\!
[\psi_1(\bm r) \ \psi_0(\bm r) \ \psi_{-1}(\bm r)]^{\rm T}$
as a sum of the condensate wave function ${\bm\Psi}(\bm r)$ 
and the quasiparticle field ${\bm\phi}(\bm r)$ as
\begin{align}
		  {\bm\psi}(\bm r) = {\bm\Psi}(\bm r) + {\bm\phi}(\bm r).
		  \label{shift}
\end{align}
We next introduce the Matsubara Green's function in the Nambu space as  
\begin{subequations}
	\begin{align}
        \hat G(\bm r, \bm r', \tau) &= 
         - \hat \tau_3 \langle T_{\tau} \hat{\bm\phi}(\bm r, \tau) \hat{\bm\phi}^\dagger(\bm r') \rangle
         \\
         & \equiv \begin{bmatrix}
                 \underline{G}(\bm r, \bm r', \tau) & \underline{F}(\bm r,\bm r',\tau) \\
                 -\underline{F}^*(\bm r,\bm r',\tau) &  
                 -\underline{G}^*(\bm r, \bm r', \tau)
         \end{bmatrix},
        \label{hatG}
	\end{align}
\end{subequations}
with $\hat \tau_3$ denoting the third Pauli matrix
in the Nambu space.
Equation (\ref{hatG}) can be Fourier-transformed as
\begin{align}
        \hat G(\bm r, \bm r',\tau) = \frac{1}{\beta}\sum_n \hat G(\bm r, \bm r',i\omega_n) e^{-i\omega_n \tau},
        \label{hatG-Matsu}
\end{align}
where $\beta \!\equiv\! T^{-1}$ and 
$\omega_n$ is the Matsubara frequency $\omega_n\!=\!{2n \pi}/{\beta}$ 
with $n=0,\pm 1, \pm 2,\cdots$.

Using eq.\ (\ref{hatG-Matsu}), we now write down 
our mean-field Luttinger-Ward functional for BEC as\cite{LW,Kita-FP}
\begin{align}
        &\Omega = \int {\rm d}\bm r \Psi^\dagger(\bm r) (H_0 - \mu)  \Psi(\bm r)
		  \notag\\
        &+ \frac{1}{2 \beta} \sum_{n= - \infty}^{\infty}{\rm Tr}
        \left[\ln(\hat K + \hat \Sigma - i\omega_n \hat{1})  + \hat G \hat \Sigma \right]
        \hat 1(\omega_n) 
		  + \Phi \, .
        \label{eq:LW-definition}
\end{align}
Here $\hat K$ is defined by $\hat K\!\equiv\!\hat \tau_3 (H_0 - \mu)$,
$\hat 1(i \omega_n)$ denotes
\begin{align}
		\hat 1(\omega_n)= 
		\begin{bmatrix}\underline{1} {\rm e}^{i\omega_n 0^+} &\underline{0} \\ 
				  \underline{0} & \underline{1} {\rm e}^{-i\omega_n 0^+} \end{bmatrix},
\end{align}
with $0^{+}$ an infinitesimal positive constant, and 
{\rm Tr} includes integration over space variables.
The quantity $\hat \Sigma$ is the proper self-energy obtained from the functional 
$\Phi = \Phi[\Psi,\Psi^*,\hat G]$ by
\begin{align}
        \hat \Sigma(\bm r, \bm r',\omega_n) 
        = -2 \beta \frac{\delta \Phi}{\delta \hat G(\bm r', \bm r,\omega_n)}.
        \label{eq:ISE}
\end{align}
With eq.\ (\ref{eq:ISE}), functional $\Omega$ is stationary with respect to a variation 
of $\hat G$ which satisfies 
Dyson's equation:
\begin{align}
        \hat G^{-1} = i \omega_n \hat{1} - \hat K - \hat \Sigma  \, .
        \label{Dyson}
\end{align}
The condensate wave function $\bm \Psi$ in equilibrium is also determined by 
$\delta \Omega/ \delta \bm \Psi^{*} =0$.
Noting $\delta \Omega/ \delta \hat G =0$, we obtain 
\begin{align}
		  (H_0-\mu) \bm \Psi (\bm r) 
		  + \frac{\delta \Phi}{\delta \bm \Psi^{*}({\mib r})} = 0 \, .
		  \label{eq:cond-WF}
\end{align}
Equations  (\ref{eq:ISE}) and (\ref{eq:cond-WF}) constitute 
the $\Phi$-derivative approximation
where various conservation laws are obeyed automatically.
 This is one of the main advantages of using the Luttinger-Ward functional.

Now, we choose $\Phi$ so that the Hugenholtz-Pines theorem is 
satisfied simultaneously; see Appendix for the
multi-component version of the Hugenholtz-Pines theorem.
Explicitly, our $\Phi$ is given by
\begin{align}
		 &\Phi =  \frac{1}{2}\sum_{\nu}\int {\rm d} \bm r {\rm d} \bm r'
		 \mathcal{U}_{\nu}(\bm r- \bm r')
		  \notag\\
        &\times \Bigl \{
        \frac{1}{2}{\rm Tr}[\hat{A}^{\nu} \hat {\bm \Psi}(\bm r)  \hat {\bm \Psi}^{\dagger}(\bm r)] 
        \frac{1}{2}{\rm Tr}[\hat{A}^{\nu} \hat {\bm \Psi}(\bm r') \hat {\bm \Psi}^{\dagger}(\bm r')] 
        \notag \\
       &- \frac{2}{\beta} \sum_n \Big(
       \frac{1}{2}{\rm Tr}[\hat{A}^{\nu}\hat {\bm \Psi}(\bm r) \hat {\bm \Psi}^\dagger(\bm r)]
			\notag \\ &\hspace{1.5cm} \times
         \frac{1}{2}{\rm Tr}[\hat{A}^{\nu} \hat \tau_3 \hat G(\bm r',\bm r',i \omega_n)
         \hat 1(\omega_n)]
         \notag\\
        & +\frac{1}{2}{\rm Tr}[\hat{A}^{\nu}\hat\tau_3\hat {\bm \Psi}(\bm r) \hat {\bm \Psi}^\dagger(\bm r')
         \hat{A}^{\nu}\hat G(\bm r',\bm r,\omega_n)\hat 1(\omega_n)]
          \Big)\notag\\
        & +\frac{1}{\beta^2}\sum_{n,m}\Big(
        \frac{1}{2}{\rm Tr}[\hat{A}^{\nu}\hat \tau_3 \hat G(\bm r,\bm r,\omega_n)]
		  \notag\\ & \hspace{1.5cm} \times
        \frac{1}{2}{\rm Tr}[\hat{A}^{\nu}\hat \tau_3 \hat G(\bm r',\bm r',\omega_m)]
        \notag\\
       & +\frac{1}{2}{\rm Tr}[\hat{A}^{\nu}\hat G(\bm r,\bm r',\omega_n)\hat 1(\omega_n)
                            \hat{A}^{\nu}\hat G(\bm r',\bm r,\omega_m)\hat 1(\omega_m)]
        \Big)
        \Bigr\}.
        \label{eq:functional-Phi}
\end{align}
This functional $\Phi$ is different from $\Phi_{\rm HFB}$ of the HFB theory
in the Fock terms, i.e., the third and the fifth terms on the right-hand side of 
eq.\ (\ref{eq:functional-Phi}).
Indeed, $\Phi_{\rm HFB}$ is recovered from $\Phi$ by 
replacing $\hat \tau_3\hat {\bm \Psi} \hat {\bm \Psi}^\dagger \rightarrow  \hat {\bm \Psi} \hat {\bm \Psi}^\dagger$ 
and $ \hat G \rightarrow \hat \tau_3 \hat G$.
Put it another way, $\Phi$ is obtained from $\Phi_{\rm HFB}$
by the following subtraction:
\begin{align}
		  \Phi =& \Phi_{\rm HFB} 
		  - \sum_{\nu} \int {\rm d} \bm r {\rm d} \bm r' \mathcal{U}_{\nu}(\bm r - \bm r')
		  \notag\\ &\times
		  \Bigr(- \frac{1}{\beta}\sum_n 
        {\rm Tr}[\hat{A}^{\nu} \hat {\bm \Psi}(\bm r) \hat{\bm \Psi}^\dagger (\bm r')
		  			  \hat{A}^{\nu}\hat \tau_3 \hat F(\bm r',\bm r,\omega_n) \hat 1(\omega_n)]
					  \notag \\
					  &+ \frac{1}{2\beta^2} \sum_{n,m}
					  {\rm Tr}[\hat{A}^{\nu}\hat \tau_3 \hat G(\bm r,\bm r',\omega_n)\hat 1(\omega_n)
					  \notag\\ &\hspace{1.5cm} \times
					  \hat{A}^{\nu} \hat \tau_3 \hat F(\bm r',\bm r,\omega_m)\hat 1(\omega_m)]
		  \Bigl) \, ,
		  \label{}
\end{align}
with
\begin{align}
		  \hat F(\bm r, \bm r',\omega_n) = \begin{bmatrix}
                \underline 0 & \underline{F}(\bm r,\bm r',\omega_n)\\
                -\underline{F}^*(\bm r, \bm r',\omega_n) & \underline 0
		  									\end{bmatrix}
		  .\label{}
\end{align}
This subtraction has the effect to cancel the overcounting of the 
anomalous pair correlation in the HFB theory, thereby leading to
a mean-field theory which satisfies the Hugenholtz-Pines theorem.

Now, the proper self-energy is obtained with eqs.\ 
(\ref{eq:ISE}) and (\ref{eq:functional-Phi}) as
\begin{align}
		  &\hat \Sigma(\bm r, \bm r')
		  = 
        \begin{bmatrix}
                \underline\Sigma(\bm r, \bm r')   &  \underline\Delta(\bm r, \bm r') \\
               -\underline\Delta^*(\bm r, \bm r') & -\underline\Sigma^*(\bm r, \bm r')
        \end{bmatrix}
        \notag\\
		   &=\sum_{\nu}\biggl\{ \frac{1}{2}\delta(\bm r - \bm r')
			\hat{A}^{\nu}\hat \tau_3 
			\int {\rm d}\bm r'' \mathcal{U}_{\nu}(\bm r- \bm r'')
			\notag\\
		   & \hspace{3.5mm}\times 
		  {\rm Tr}[\hat{A}^{\nu}\hat \tau_3 \hat \rho(\bm r'',\bm r'')] +
			\mathcal{U}_{\nu}(\bm r - \bm r') \hat{A}^{\nu} 
			\hat \rho(\bm r,\bm r')\hat{A}^{\nu}\biggr\} ,
		  \label{hatSigma}
\end{align}
where $\hat{\rho}$ is the one-particle density matrix in the Nambu space
defined by
\begin{align}
		  &\hat \rho(\bm r, \bm r') =
        \begin{bmatrix}
                \underline\rho(\bm r, \bm r')   &  \underline{\tilde\rho}(\bm r, \bm r') \\
               -\underline{\tilde\rho}^*(\bm r, \bm r') & -\underline\rho^*(\bm r, \bm r')
        \end{bmatrix}
        \notag\\
        &  =\hat \tau_3 \vec\Psi(\bm r)\vec\Psi^\dagger(\bm r')
        +\hat{\rho}^{({\rm qp})}(\bm r, \bm r') \, ,
		  \label{hatRho}
\end{align}
with
\begin{align}
		  \hat{\rho}^{({\rm qp})}(\bm r, \bm r')\equiv
		  -\frac{1}{\beta}\sum_n \hat G(\bm r, \bm r', \omega_n)\hat 1(\omega_n)\, .
		  \label{rho^qp}
\end{align}
The submatrices $\underline\Sigma$ and $\underline\Delta$
satisfy
$\underline{\Sigma}^*(\bm r,\bm r')\!=\! \underline{\Sigma}^{\rm T}(\bm r',\bm r)$ and
$\underline{\Delta}(\bm r,\bm r')=\underline{\Delta}^{\rm T}(\bm r',\bm r)$.
Similarly, we have 
$\underline{\rho}^*(\bm r,\bm r')\!=\! \underline{\rho}^{\rm T}(\bm r',\bm r)$ and
$\underline{\tilde\rho}^{\rm T}(\bm r,\bm r')=\underline{\tilde\rho}(\bm r',\bm r)$.

Next, eqs.\ (\ref{eq:cond-WF}) and (\ref{eq:functional-Phi}) lead 
to the the equation for the condensate wave function:
\begin{align}
		  &(H_0 - \mu ) \bm\Psi(\bm r) 
		  \notag\\ 
        &\hspace{5mm}
		  + \int {\rm d}\bm r'[\underline\Sigma(\bm r,\bm r')\bm\Psi(\bm r) 
        - \underline\Delta(\bm r, \bm r')\bm\Psi^*(\bm r')]
		  = \underline 0 \, .
		  \label{eq:ex-GP}
\end{align}
Equation (\ref{eq:ex-GP}) is the GP equation 
in our mean-field theory. 
In the homogeneous case, we can write
$\bm \Psi \!\rightarrow\! \sqrt{n_0} {\bm\eta}$, 
where $n_0$ is the condensate density and
${\bm\eta}$ ($|{\bm\eta}|\!=\!1 $) 
specifies the direction of the
order parameter.
Equation (\ref{eq:ex-GP}) then reduces to
$\mu \bm\zeta \!=\! 
\underline{\Sigma}_{\bm k=\bm 0} \bm\zeta 
- \underline{\Delta}_{\bm k=\bm 0} \bm\zeta^*$,
which is the Hugenholtz-Pines relation of the multi-component case;
see eq.\ (\ref{HP-multi}) in Appendix.
It tells us the existence of a single gapless excitation spectrum
in the system, as shown explicitly below eq.\ (\ref{eq:BdG-k}).
Note that $\mu$ and ${\bm\eta}$ should be determined here as the lowest 
eigenvalue and its eigenstate, respectively.
Finally, the thermodynamic relation 
$N \!=\! - \partial \Omega/ \partial \mu$ 
yields the expression for the total particle number as
$N \!=\! {\rm Tr}\int \underline{\rho}(\bm r, \bm r){\rm d}{\bm r}$.

In order to diagonalize Green's function which satisfies eq.\ (\ref{Dyson}),
we consider the Bogoliubov-de Gennes (BdG) equation:
\begin{align}
        \int {\rm d}\bm r' \hat H(\bm r, \bm r') 
        \begin{bmatrix} {\bm u}_{j} (\bm r') \\ -{\bm v}_{j}^*(\bm r')\end{bmatrix}
               = 
               E_{j} \begin{bmatrix} {\bm u}_{j} (\bm r) \\ -{\bm v}_{j}^* (\bm r)\end{bmatrix},
        \label{eq:BdG}
\end{align}
where 
$\hat H(\bm r, \bm r') \!\equiv\! \hat K (\bm r) \delta(\bm r - \bm r') + \hat \Sigma(\bm r, \bm r')$.
We assume that the eigenstate for $E_{j} >0$
can be normalized as
\begin{align}
        \langle \bm u_{j} |\bm u_{j'} \rangle - \langle \bm v_{j} |\bm v_{j'} 
        \rangle^* = \delta_{jj'}
		  .\label{}
\end{align}
It then follows from the symmetry of $\hat{H}$ that 
the eigenvector with the eigenvalue $-E_{j}$ is given by 
$[-{\bm v}_{j}^{\rm T}, \bm u_{j}^{\dagger }]^{\rm T}$. \cite{Kita-FP}
Let us introduce matrices $\hat U$ and $\hat E$ by
\begin{align}
        \hat U \equiv \begin{bmatrix} \underline{U} & -\underline{V} \\
                -\underline{V}^* & \underline{U}^* \end{bmatrix},
                \hspace{5mm}
                \hat E \equiv \begin{bmatrix} \underline{E} & \underline{0} \\
                \underline{0} & -\underline{E} \end{bmatrix},
        \label{}
\end{align}
with
$\underline{U} \!\equiv\! \bigl({\bm u}_{j_{1}}\,{\bm u}_{j_{2}}\,
\cdots\bigr) $,
$\underline{V} \!\equiv\! \bigl({\bm v}_{j_{1}}\,{\bm v}_{j_{2}}\,
\cdots\bigr)$ and $(\underline{E})_{jj'}=E_{j}\delta_{jj'}$.
Using these matrices, the BdG equation can be written compactly as
$ \int {\rm d} \bm r'\hat H(\bm r , {\bm r'})\hat U({\bm r'}) = \hat U(\bm r) \hat E$.
It hence follows that we can expand Green's function as
\begin{align}
        \hat G(\bm r, \bm r',\omega_n) = \hat U(\bm r) 
        (i \omega_n \hat 1 - \hat E)^{-1} \hat U^{\dagger}(\bm r').
		  \label{}
\end{align}
We now carry out the summation over the Matsubara frequency in eq.\ (\ref{rho^qp}).
We then obtain the quasiparticle contribution 
to the density matrix in eq.\ (\ref{hatRho}) as
\begin{subequations}
\begin{align}
        &\underline{\rho}^{({\rm qp})}(\bm r,\bm r') 
        = \langle T_{\tau} \phi(\bm r,0^-) \phi^\dagger(\bm r')\rangle
		  \notag \\
        &= \left\{ \underline{U}(\bm r) f(\underline{E}) 
        \underline{U}^\dagger (\bm r') 
        + \underline{V}(\bm r)[\underline 1 + f(\underline{E})] 
        \underline{V}^\dagger(\bm r')
        \right\},
\end{align}
\begin{align}
		  &\underline{\tilde\rho}^{({\rm qp})}(\bm r,\bm r') 
		  = \langle \phi(\bm r) \phi(\bm r')\rangle
		  \notag\\
		  &= \frac{1}{2}\Big\{\underline{V}(\bm r) \left[\underline {1} + 
		  2 f(\underline{E})\right] \underline{U}^{\rm T}(\bm r') 
		  \notag\\ &\hspace{6mm}
		        + \underline{U}(\bm r) \left[\underline 1 + 2 f(\underline{E})\right] 
		        \underline{V}^{\rm T}(\bm r') \Big\},
\end{align}
\label{eq:explicit-GF}
\end{subequations}

\noindent
where $f$ is the Bose distribution function.

Equations (\ref{eq:ex-GP}) and (\ref{eq:BdG}) with eqs.\ (\ref{hatSigma}), (\ref{hatRho})
and (\ref{eq:explicit-GF})
form a closed set of self-consistent equations
to determine the equilibrium.

Using eqs.\ (\ref{eq:ex-GP}) and (\ref{eq:BdG}),
we can transform the thermodynamic potential of eq.\ (\ref{eq:LW-definition}) 
into
\begin{align}
        &\Omega_{\rm eq} = \beta^{-1} \sum_{j} \ln(1 - {\rm e}^{-\beta E_{j}}) 
		  - \sum_j E_j \int  |{\bm v}_j(\bm r)|^2 \,{\rm d} \bm r
		  \notag\\
		  &- \frac{1}{2} \int {\rm d} \bm r {\rm d}\bm r'  
		  {\rm Tr}\left[ \underline{\Sigma}(\bm r, \bm r')\underline{\rho}(\bm r',\bm r) 
		  - \underline{\Delta}(\bm r, \bm r')\underline{\tilde\rho}(\bm r',\bm r)\right] .
		  \label{}
\end{align}
Also, entropy is obtained as
\begin{align}
		  S = - \sum_{j} \left\{ f(E_{j})\ln f(E_{j}) - [1+f(E_{j})] \ln [1 + f(E_{j})]\right\} .
		  \label{}
\end{align}
Finally, the specific heat is calculated by the thermodynamic relation 
$C = - T \partial S/\partial T$.

\section{Uniform Gas under Constant Density}
\subsection{Equations to determine equilibrium}

We now apply the previous formulation to a uniform Bose gas with
fixed density $n \equiv N/\mathcal{V}$, where
$\mathcal{V}$ denotes volume of the system.
In this case, the condensate wave function $\bm \Psi$ can be written as
\begin{align}
		  \bm \Psi = \sqrt{n_0} {\bm\eta},
		  \label{Psi-uniform}
\end{align}
with $n_0\!\equiv\!N_{0}/{\cal V}$ the condensate density
and ${\bm\eta}$ a constant vector with $|{\bm\eta}|\!=\! 1$.
Also, eq.\ (\ref{hatRho}) may be expanded in plane waves as
\begin{align}
		  \hat{\rho}({\bm r},{\bm r}')=\frac{1}{\mathcal{V}} 
		  \sum_{\bm k} 
		  \begin{bmatrix}
		  \underline{\rho}({\bm k}) & \underline{\tilde{\rho}}({\bm k})\\
		  -\underline{\tilde{\rho}}(-{\bm k}) & -\underline{\rho}(-{\bm k})
		  \end{bmatrix}
        {\rm e}^{i \bm k \cdot ({\bm r}-{\bm r}')},
\label{rho-uniform}
\end{align}
with
\begin{subequations}
\label{rho_k-tot}
\begin{align}
		  \underline{\rho}({\bm k}) = 
		  \delta_{{\bm k}{\bf 0}}N_{0}{\bm\eta}{\bm\eta}^{\dagger}
		  +\underline{\rho}^{({\rm qp})}({\bm k}),
		  \label{rho_k}
\end{align}
\begin{align}
		  \underline{\tilde\rho}({\bm k}) = 
		  \delta_{{\bm k}{\bf 0}}N_{0}{\bm\eta}{\bm\eta}^{\rm T}
		  +\underline{\tilde\rho}^{({\rm qp})}({\bm k}) .
\end{align}
\end{subequations}
The transition temperature of the noninteracting three-component BEC
is given by
\begin{align}
        T_0 = \frac{2 \pi}{M} \left[\frac{n}{3 \zeta(\frac{3}{2})}\right]^{\frac{2}{3}},
		  \label{}
\end{align}
where $\zeta(\frac{3}{2}) \simeq 2.612$ is the Riemann $\zeta$-function.
This $T_{0}$ is $3^{-2/3}$ times the transition temperature 
of the one-component BEC.

The self-energy (\ref{hatSigma}) 
becomes $\bm k$-independent in momentum space 
for the contact interaction of eq.\ (\ref{U_nu}).
Indeed, $\underline{\Sigma}$ and
$\underline{\Delta}$ in $\bm k$ space are obtained as
\begin{subequations}
\begin{align}
        \frac{\underline{\Sigma}}{T_0} =
        \frac{2}{N}\left[3 \zeta\!\left(\frac{3}{2}\right)\right]^{2/3}
        \sum_{\nu\bm k}\delta_{\nu} 
       \bigl[\underline{A}^{\nu}{\rm Tr}\underline{\rho}({\bm k})\underline{A}^{\nu}
       + \underline{A}^{\nu}\underline{\rho}({\bm k})\underline{A}^{\nu}\bigr],
		 \label{eq:uni-sigma}
\end{align}
\begin{align}
		  \frac{\underline \Delta}{T_0} 
		  =\frac{2}{N}\left[3 \zeta\!\left(\frac{3}{2}\right)\right]^{2/3}
        \sum_{\nu\bm k}\delta_{\nu} 
		 \underline{A}^{\nu} \underline{\tilde{\rho}}({\bm k}) 
		  (\underline{A}^{\nu})^{\rm T} ,
		  \label{eq:uni-delta}
\end{align}
\label{eq:uniform-SE}
\end{subequations}

\noindent
where $\delta_{\nu}$ denotes dimensionless interaction strength
defined in terms of eq.\ (\ref{c_0c_2}) by
\begin{align}
		  \delta_{\nu} = \left\{
		  \begin{array}{ll}
		  \vspace{2mm}
		  \displaystyle\frac{M}{4 \pi } c_{0} n^{1/3} & :\nu=0 \\
		  \displaystyle\frac{M}{4 \pi } c_{2} n^{1/3} & :\nu=1,2,3 
		  \end{array}
		  \right. .
		  \label{delta_nu}
\end{align}
The GP equation (\ref{eq:ex-GP}) is transformed 
by using eq.\ (\ref{Psi-uniform}) into
\begin{align}
		  \mu {\bm\eta} = \underline \Sigma {\bm\eta} - \underline \Delta {\bm\eta}^* .
		  \label{eq:HP}
\end{align}
This is the Hugenholtz-Pines relation of the three-component system;
see Appendix for details.
The BdG equation (\ref{eq:BdG}) now reads
\begin{align}
		  \hat H_{\bm k} 
		  		\begin{bmatrix} \bm u_{\bm k \sigma} \\ 
		  		- \bm v_{\bm k \sigma}^{*} \end{bmatrix}
						  = E_{\bm k \sigma}
		  		\begin{bmatrix} \bm u_{\bm k \sigma} \\ 
		  		- \bm v_{\bm k \sigma}^{*} \end{bmatrix},
		  \label{eq:BdG-k}
\end{align}
where 
$\hat H_{\bm k} \equiv \hat \tau_3 \left( \frac{k^2}{2M} - \mu \right) + \hat \Sigma$,
the eigenstate is normalized as $|u_{\bm k \sigma}|^{2}\!-\!|v_{\bm k \sigma}|^2\!=\! 1$, 
and the subscript $\sigma\!=\!0,\pm 1$ distinguishes the three eigenvalues
of each ${\bm k}$.
This eigenvalue problem includes eq.\ (\ref{eq:HP}) as the special case for
$E_{{\bm k} = {\bf 0}\sigma}\!=\!0$ with $u_{\bm k= {\bf 0} \sigma}\!=\!
v_{\bm k= {\bf 0} \sigma}\!=\!\bm\zeta$.
It hence follows that the quasiparticle excitation is gapless
as ${\bm k}\!\rightarrow\!{\bf 0}$ at least for one branch among
$\sigma\!=\!0,\pm 1$.
The quasiparticle contribution (\ref{eq:explicit-GF}) to the density matrix 
is now given by
\begin{subequations}
\begin{align}
		  \underline \rho^{({\rm qp})}(\bm k) = \left\{\underline U(\bm k) f(\underline E_{\bm k}) \underline U^\dagger(\bm k)
		   \!+\!\underline V(\bm k) [\underline 1 \!+\! f(\underline E_{\bm k})] \underline V^\dagger(\bm k)
					\right\},
					\label{eq:uni-G}
\end{align}
\begin{align}
		  \underline{\tilde\rho}^{({\rm qp})}(\bm k) = \frac{1}{2}
		  \bigl\{ \underline V(\bm k)[\underline 1 + 2f(\underline E_{\bm k})] \underline U^{\rm T}(\bm k)
		  \notag\\
		  + \underline U(\bm k) [\underline 1 + 2f(\underline E_{\bm k})] \underline V^{\rm T}(\bm k)
					\bigr\},
\label{eq:uni-F}
\end{align}
\label{eq:uni-GF}
\end{subequations}

\noindent
with
$ \underline{U} (\bm k) \!\equiv\! [\bm u_{\bm k 1} \ \bm u_{\bm k 0} \ \bm u_{\bm k -1}]$ and
$ \underline{V} (\bm k) \!\equiv\! [\bm v_{\bm k 1} \ \bm v_{\bm k 0} \ \bm v_{\bm k -1}]$.
Finally, we sum eq.\ (\ref{rho_k}) over ${\bm k}$ and take trace of it 
subsequently.
We thereby obtain an expression for the condensate density as
\begin{align}
		  \frac{n_0}{n} = 1 - \frac{1}{N}\sum_{\bm k} {\rm Tr}
		  \underline{\rho}^{({\rm qp})}(\bm k).
		  \label{eq:cond-density}
\end{align}

\subsection{Numerical procedures}

Numerical calculations of the above self-consistent equations 
have been carried out as follows.

(i) The chemical potential $\mu$ and the vector ${\bm \eta}$
are determined by eq.\ (\ref{eq:HP}) 
as the smallest eigenvalue and its eigenstate.
To carry this out, we have to know the expressions of $n_{0}$,
$\underline{\rho}^{({\rm qp})}({\bm k})$ and 
$\underline{\tilde\rho}^{({\rm qp})}({\bm k})$ in advance,
as seen from eqs.\ (\ref{rho_k-tot}) and (\ref{eq:uniform-SE}).
We use the values from the previous calculation for them;
we initially set $n_{0}\!=\!n$ and
$\underline{\rho}^{({\rm qp})}\!=\! \underline{\tilde\rho}^{({\rm qp})} 
\!=\! \underline 0$ to start the whole calculation at $T\!=\!0$.
We determine ${\bm\eta}$ self-consistently
so that it actually corresponds to the eigenstate of the smallest eigenvalue.

(ii) Using $\mu$ and $\hat \Sigma$ thus calculated, 
we solve the BdG equation (\ref{eq:BdG-k}) to
obtain the quasi-particle spectrum $E_{{\bm k}\sigma}$ and 
the matrix ${\hat U}_{{\bm k}}$.

(iii) With $E_{{\bm k}\sigma}$ and ${\hat U}_{{\bm k}}$, 
we compute 
$\underline{\rho}^{({\rm qp})}({\bm k})$,
$\underline{\tilde\rho}^{({\rm qp})}({\bm k})$ and $n_{0}$ by
eqs.\ (\ref{eq:uni-G}), (\ref{eq:uni-F}) and (\ref{eq:cond-density}), respectively.

We iterate procedure (i)-(iii) until a convergence is reached.
We start the whole calculation from $T\!=\!0$ and increase the temperature
gradually. 
We identify $T_{c}$ numerically as the point where we can no longer find
a solution of positive $n_{0}$.

When we calculate $\underline \Delta$ for $T\!<\!T_c$
by using eqs.\ (\ref{eq:uni-delta}) and (\ref{eq:uni-F}),
we observe that the summation of 
$\underline{\tilde \rho}^{\rm (qp)}(\bm k)$
over $\bm k$ diverges. In order to remove it numerically,
we have introduced the energy cutoff $\epsilon_c$ so that it
satisfies 
$1 \!\ll\! \epsilon_c \!\ll\! 0.3 \delta_0^{-2}$.\cite{Kita-FP}

For $T > T_c$, we put $n_0 = 0$, 
$\underline \Delta \!=\! \underline{0}$ and $\underline F \!=\! \underline 0$.
The calculation procedures (i)-(iii) are nearly the same except that
the chemical potential is now determined by
\begin{align}
        N = \sum_{\bm k \sigma} \frac{1}{\exp[\beta E_{\bm k \sigma}(\mu) ] - 1}.
		  \label{}
\end{align}

\section{Numerical Results}
\subsection{Thermodynamic properties}

We have fixed the interaction parameters of eq.\
(\ref{delta_nu}) as $\delta_0\!=\! 0.0075$ and
$c_2/c_0 \!=\! 0$, $\pm 0.1$. 
We have confirmed that the stable state
is given by $\bm \eta \!=\![0 1 0]^{\rm T}$ ($\bm \eta \!=\![1 0 0]^{\rm T}$)
at all temperatures for the antiferromagnetic (ferromagnetic) interaction
of $c_2/c_0 \!=\! 0.1$ ($c_2/c_0 \!=\! - 0.1$).
We have found for all these cases that the transition temperature 
$T_{c}$ is increased over $T_0$
by $(T_c \!-\! T_0)/T_0\!\approx\!2 \delta_0$,
which is of the same order as 
$(T_c \!-\! T_0)/T_0 \!\simeq\! 2.33 \delta$ 
of the single-component case.\cite{Kita-LT,Kita-FP}
We have also observed that 
$T_{c}$ increases
monotonically as a function of $c_{2}/c_{0}$.

\begin{figure}[t]
        \begin{center}
                \includegraphics[width=7cm,clip]{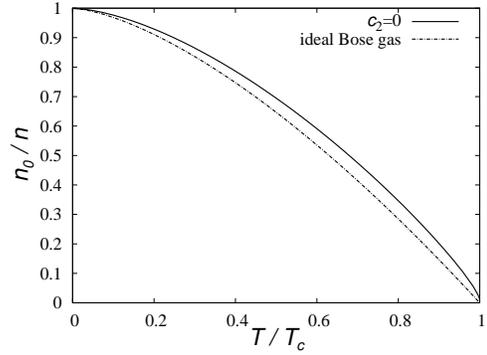}
        \end{center}
        \caption{Temperature dependence of density of particle.}
        \label{fig:density}
\end{figure}

Figure \ref{fig:density} shows temperature dependence of
the condensate density in comparison with the ideal Bose gas
result $n_0/n \!=\!1 \!-\! (T/T_c)^{3/2}$.
We observe that $n_0$ becomes larger as the interaction is increased.
We have also confirmed that $n_0$ depends little on $c_{2}$ 
as long as $|c_2/c_0|$ is much smaller than $1$.

\begin{figure}[t]
        \begin{center}
                \includegraphics[width=7cm,clip]{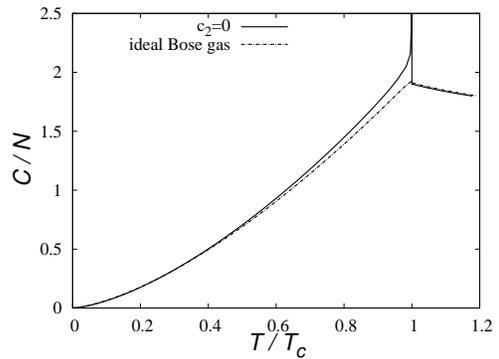}
        \end{center}
        \caption{Temperature dependence of specific heat.}
        \label{fig:specific_heat}
\end{figure}

Figure \ref{fig:specific_heat} displays 
temperature dependence of the specific heat per particle in comparison with
the ideal gas result. 
As seen clearly, a divergent behavior shows up
just below $T_{c}$ for the finite interaction.
It is quite similar to the behavior of the single-component system.

\begin{figure}[t]
		  \begin{center}
					 \includegraphics[width=7cm,clip]{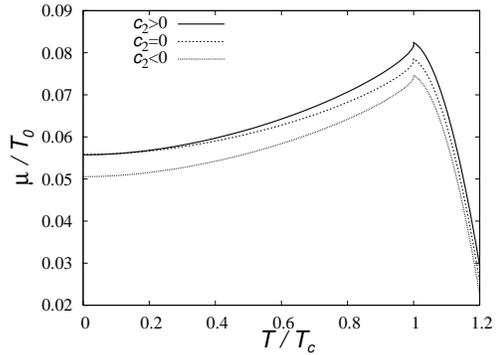}
		  \end{center}
		  \caption{chemical potential}
		  \label{fig:chem_3}
\end{figure}

Figure\ \ref{fig:chem_3} plots temperature dependence
of the chemical potential scaled by $T_0$.
With finite interaction, we find a small peak 
with a discontinuity at $T\!=\!T_c$.
This discontinuity is a signal of a first-order transition
and also seen in the single-component case.

\subsection{Energy dispersion of antiferromagnetic interaction}
We next study the energy dispersion of the antiferromagnetic interaction 
($c_2/c_0\!=\!0.1$) at $T\!=\!0$
and compare our result with the prediction of the Bogoliubov approximation (BA).
In BA, the quantities $\mu$, ${\bm \eta}$ and $E_{{\bm k}\sigma}$
can all be obtained analytically as $\mu \! = \! c_0 n_0$,
$\bm \eta \!=\! [0 1 0]^{\rm T}$, 
$E_{\bm k 0} \!=\! \sqrt{\epsilon_{\bm k}(\epsilon_{\bm k} \!+\! 2 c_0 n_0)}$
and
$E_{\bm k \pm 1} \!=\! \sqrt{\epsilon_{\bm k}(\epsilon_{\bm k} \!+\! 2 c_2 n_0)}$
with $\epsilon_{\bm k} \!=\! {\bm k^2}/{2M}$; thus,
$E_{\bm k \pm1}$ is degenerate in BA. 

\begin{figure}[t]
		  \begin{center}
					 \includegraphics[width=7cm,clip]{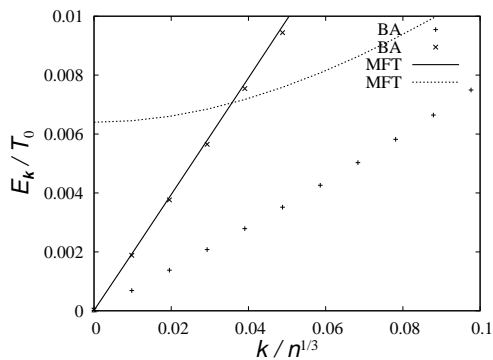}
		  \end{center}
		  \caption{The energy dispersion of antiferromagnetic interaction at $T=0$}
		  \label{fig:dispt0_af}
\end{figure}

Figure \ref{fig:dispt0_af} shows energy dispersion at $T \!= \!0$.
The BA predicts that all the three branches have $k$-linear 
sound-wave dispersions in the long wavelength limit and
two of them are degenerate.
Our theory also yields a single gapless excitation 
in agreement with the Hugenholtz-Pines theorem,
and its sound velocity $c_{\rm AF}$ coincides almost completely with
the expression ${c_0 n_0}/{M}$  of BA for the non-degenerate
dispersion, as seen in Fig.\ \ref{fig:dispt0_af}.
As for the other two branches, 
our mean-field theory predicts degenerate gapful excitations.
This gap is brought about by the anomalous pair potential
of the quasiparticle field in the BdG equation,
which is absent in BA.

\begin{figure}[t]
		  \begin{center}
					 \includegraphics[width=7cm,clip]{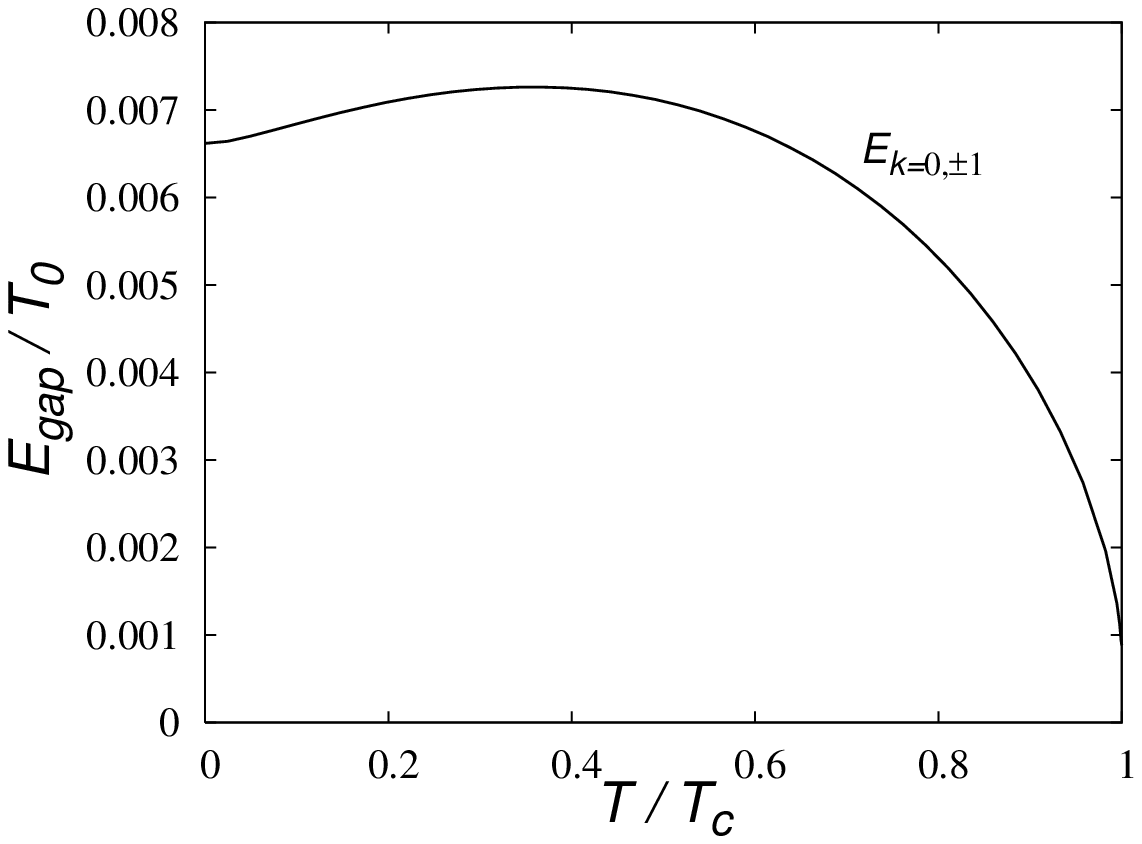}
		  \end{center}
		  \caption{Temperature dependence of excitation energy gap in anti-ferro interaction.}
		  \label{fig:energygap_af}
\end{figure}

The qualitative features of the energy dispersions mentioned above remain invariant
up to $T_{c}$.
The sound velocity decreases monotonically towards $T_{c}$.
In contrast, the energy gap between $E_{{\bm k}={\bm 0},0}$ and 
$E_{{\bm k}={\bm 0},\pm 1}$ displays a more complicated behavior.
Figure \ref{fig:energygap_af} shows temperature dependence of
the energy gap.
The gap has a local maximum as a function of temperature,
which may be traced to the temperature dependence of the anomalous 
pair correlation
$\sum_{\bm k} \underline{\tilde \rho}(\bm k)^{(\rm qp)}$.
Indeed, $\sum_{\bm k} \underline{\tilde \rho}(\bm k)^{(\rm qp)}$
has a local maximum even in the single-component case,
which is due to the competition between decreasing 
$\underline \Delta$ and increasing Bose distribution function
as $T$ is raised.

We finally comment on the finite energy gap of $E_{{\bm k}={\bf 0},\pm 1}$
predicted by our mean-field theory in connection with the collective 
excitations. 
Gavoret and Nozi\`eres\cite{GN64} performed a detailed study 
on the structure of the perturbation expansion
for the one-component BEC at $T\!=\! 0$.
They have thereby established that 
the single-particle (i.e., Bogoliubov) and collective excitations 
share common dispersions.
Indeed, the statement holds even at any finite temperatures of 
$T\!\leq\!T_{c}$, which stems from the fact
that the single-particle propagator appears in the intermediate state
of the collective propagator in the 
condensate; see eq.\ (2.33) 
of Sz\'epfalusy and Kondor\cite{SK74}
and also their discussion in the paragraph below 
eq.\ (2.34).
However, it should be noted that 
this theorem only states that the single-particle excitation
spectra are also included in the collective excitations,
but not vice versa.
Thus, it is neither surprising nor contradictory that the
collective excitations have extra branches not present in the
single-particle excitations.

The theorem may be extended to multi-component systems
to tell us that all the single-particle excitation
spectra are also present in the collective excitations.
It is also worth noting that the Hugenholtz-Pines theorem
of the multi-component case only states the existence
of a single gapless one-particle excitation spectrum,
as shown in Appendix. Thus, the qualitative features of the
three branches $E_{{\bm k}0}$ and $E_{{\bm k}\pm 1}$ obtained here
satisfy all the exact statements,
and it is not contradictory that the collective spin-wave modes are
not included in the single-particle excitations.

Recently, Sz\'epfalusy and Szirmai\cite{dielectric}
have extended the consideration of 
Sz\'epfalusy and Kondor\cite{SK74} to the spin-$1$ system.
They conclude that there appear additional 
collective excitations due to the spin-degrees of freedom
whose spectra in the long-wavelength limit 
are also gapless and identical with those of the corresponding 
single-particle excitations. 
This result is clearly in contradiction with the finite energy
gap of $E_{{\bm k}={\bf 0},\pm 1}$ in our mean-field theory.
We suspect that the conclusion of Sz\'epfalusy and Szirmai\cite{dielectric}
on the existence of the spin-wave mode in the single-particle spectra
is due to an inappropriate assumption on the poles and zeros
of the single-particle Green's function expressed with 
the same denominator as the collective propagators.
Note also that they only confirm
their general theory within the Bogoliubov approximation
and the random-phase approximation of retaining only the Hartree term
where the anomalous pair potential due to quasiparticles 
was neglected completely.

\subsection{Energy dispersion of ferromagnetic interaction}
We move on to
study the energy dispersion of the ferromagnetic 
interaction ($c_2/c_0\!=\!-0.1$).
In this case, BA near $T\!=\!0$ predicts $\mu \!=\!(c_0+c_2)n_0$,
$\bm \eta \!=\! [100]^{\rm T}$, 
$E_{\bm k +1}\!=\!\sqrt{\epsilon_{\bm k}[\epsilon_{\bm k} \!+\! 2 (c_0 \!+\! c_2)n_0]}$,
$E_{\bm k 0}\!=\!\epsilon_{\bm k}$ and
$E_{\bm k -1}\!=\!\epsilon_{\bm k} \!-\!2 c_2 n_0$. 

\begin{figure}[t]
		  \begin{center}
					 \includegraphics[width=7cm,clip]{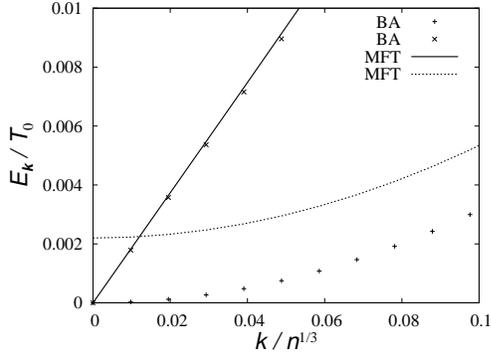}
		  \end{center}
		  \caption{Energy dispersion of ferromagnetic interaction at $T=0$}
		  \label{fig:dispt0_fe}
\end{figure}

Figure \ref{fig:dispt0_fe} shows the energy dispersions of
$E_{\bm k+1}$ and $E_{\bm k 0}$ at $T\!=\!0$.
According to BA, the two branches are both gapless and $E_{\bm k+1}$ $(E_{\bm k0})$ 
has a linear (quadratic) dispersion in the long wavelength limit.
However, our mean-field theory predicts a gapless excitation only for $E_{\bm k+1}$,
and the other branch $E_{\bm k0}$ acquires an energy gap.
The existence of a gapless excitation $E_{\bm k+1}$ agrees with the 
Hugenholtz-Pines theorem, and its sound velocity $c_{\rm FE}$ coincides 
almost completely with the expression ${(c_0 + c_2)n_0}/{M}$
of BA. On the other hand, a finite gap in $E_{\bm k0}$ is brought about by
the anomalous pair potential of the quasiparticle field.

\begin{figure}[t]
		  \begin{center}
					 \includegraphics[width=7cm,clip]{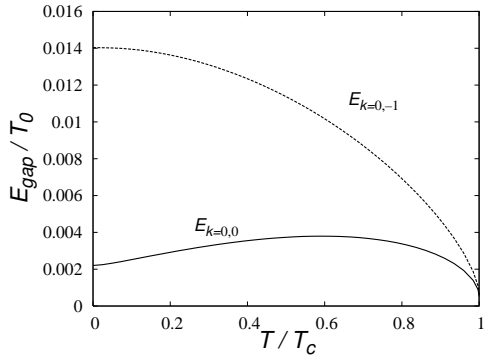}
		  \end{center}
		  \caption{Temperature dependence of excitation energy gap in ferro interaction.}
		  \label{fig:energygap_fe}
\end{figure}

We have confirmed that the vector $\bm \eta$ remains invariant
at finite temperatures. Also, the qualitative features of the dispersions do not 
change up to $T\!=\!T_c$.
The sound velocity decreases monotonically towards $T_c$.
Figure \ref{fig:energygap_fe} shows  
the energy gaps of $E_{\bm k0}$ and $E_{\bm k -1}$ at $k=0$ as a
function of temperature.
The gap of $E_{\bm k0}$ has a maximum at finite 
temperature, which has the same origin as in the 
antiferromagnetic case. 

\section{Summary}
We have extended a conserving gapless mean-field theory 
of single-component Bose-Einstein condensates
to a three-component system.
The equations to determine the equilibrium are given by
eqs.\ (\ref{eq:ex-GP}) and (\ref{eq:BdG}) with eqs.\ (\ref{hatSigma}),
(\ref{hatRho})
and (\ref{eq:explicit-GF}), which are applicable to 
nonuniform systems such as the trapped BEC with and without vortices.
They are compatible with the Hugenholtz-Pines theorem
and may be extended easily to
general multi-component systems.
Another advantage of our theory is that various conservation laws are
automatically obeyed when applied to dynamical phenomena.
Thus, the static and dynamical properties of BEC
may be treated systematically
within a single theoretical framework.

We have applied the mean-field theory to a uniform gas with constant density.
A qualitative difference is found in the quasiparticle dispersions 
between our theory and the Bogoliubov theory near $T\!=\!0$.
The Bogoliubov approximation predicts that there are three (two) gapless excitations
for the antiferromagnetic (ferromagnetic) interaction.
On the other hand, our theory yields only a single gapless branch
with a sound-wave dispersion at low energy, and the others are pushed up to 
higher energies due to the self-consistent 
potential of the quasiparticle field.
We have presented at the end of \S 4.2
a detailed discussion on those single-particle excitations
in connection with the collective excitations.
The speed of sound obtained here well agrees with the
corresponding branch in the Bogoliubov approximation.
We have also clarified finite-temperature effects
on the quasiparticle dispersions as well as temperature dependences
of basic thermodynamic quantities.

\acknowledgement
This work is supported by the 21st century COE program
``Topological Science and  Technology,'' Hokkaido University.

\appendix
\section{Hugenholtz-Pines theorem for spin-$1$ BEC}

We here generalize the Hugenholtz-Pines theorem\cite{Hugenholtz}
to the spin-$1$ BEC
by closely following the procedure of Popov.\cite{Popov87}
It will be shown that there is
at least a single gapless excitation spectrum in 
the spin-$1$ BEC (and also in any multi-component BEC).

First of all, we present 
Feynman rules for the bare perturbation expansion of
the spin-$1$ condensate as the basis for our proof.
They are obtained by slightly modifying the rules
of the single-component system.
To see this, let us rewrite the interaction of eq.\ (\ref{H'})
as
\begin{align}
		  \mathcal{H}'= \frac{1}{2} \sum_{\nu=0 }^{3}
		  & \int {\rm d} \bm r {\rm d}\bm r' \mathcal{U}_{\nu}(\bm r - \bm r') 
		  \notag\\
		  & \times \mathcal{N}
		  [{\bm\psi}^\dagger(\bm r)   \underline{A}^{\nu} 
		  {\bm\psi}(\bm r)
		   {\bm\psi}^\dagger(\bm r')  \underline{A}^{\nu} 
		   {\bm\psi}(\bm r')] \, ,
        \label{H'-App}
\end{align}
where $\underline{A}^{\nu}$ is given by eq.\ (\ref{A_nu}),
and the field operators are defined by
\begin{align}
{\bm\psi}^{\dagger} = 
        [\psi_1^\dagger \ \psi_0^\dagger \ \psi_{-1}^\dagger]\, , \hspace{5mm}
        {\bm\psi} = 
        \begin{bmatrix}
        \psi_1 \\ \psi_0 \\ \psi_{-1}
        \end{bmatrix} .
        \label{eq:Nambu-App}
\end{align}
The expression (\ref{H'-App}) 
enables us to perform the perturbation expansion
of the spin-$1$ system in terms of the Feynman diagrams of the
single-component system, i.e., no new diagrams are necessary.
To be more specific, we first consider the normal system
without the condensation.
We then notice that:
(i) the pairs
${\bm\psi}^{\dagger}({\bm r})\underline{A}^{\nu}
{\bm\psi}({\bm r})$ 
and
${\bm\psi}^{\dagger}({\bm r}')\underline{A}^{\nu}
{\bm\psi}({\bm r}')$ 
can be moved around anywhere within the ${\cal N}$ and/or $T_{\tau}$
operators in the perturbation expansion;
and (ii) a contraction of ${\bm\psi}(i)$ 
($i\!\equiv\!{\bm r}_{i}\tau_{i}$)
with ${\bm\psi}^{\dagger}(j)$ 
automatically yields the $3\times 3$ matrix 
$\langle T_{\tau}{\bm\psi}(i){\bm\psi}^{\dagger}(j)\rangle_{0}=-
\underline{G}_{0}(i,j)$,
where the subscript $0$ denotes the average 
with respect to the non-interacting
Hamiltonian.
Also, the final contraction within a closed particle 
loop can be transformed 
as $\langle T_{\tau} {\bm\psi}^{\dagger}(i)
\underline{\cal M}(i,j){\bm\psi}(j)\rangle_{0}
\!=\!{\rm Tr}\underline{\cal M}(i,j)\langle T_{\tau} {\bm\psi}(j)
{\bm\psi}^{\dagger}(i)\rangle_{0}$ with $\underline{\cal M}(i,j)$ denoting
some matrix product of contractions.
We now realize that the expansion can be performed 
in terms of the Feynman diagrams of the single-component system 
with the following modifications:
(i) Associate $\underline{G}_{0}(i,j)$
for the particle line from $j$ to $i$ and the matrix 
$\underline{A}^{\nu}$ for each vertex, and
(ii) take Tr for every closed particle line and
perform summations over $\nu$.

We next consider the condensed system and 
adopt the shift transformation (\ref{shift})
with the self-consistency condition:
\begin{equation}
\langle {\bm\phi}({\bm r})\rangle ={\bm 0}.
\label{sub-cond}
\end{equation}
Let us substitute eq.\ (\ref{shift}) into eq.\ (\ref{eq:hamiltonian})
and carry out the perturbation expansion by choosing
\begin{align}
        &\mathcal H_{0} = 
        \int {\rm d} \bm r \,{\bm\phi}^\dagger(\bm r) 
        (H_0 - \mu) {\bm\phi}(\bm r)
\end{align}
as the free field.
Once again, we only need the Feynman diagrams 
of the single-component system, and
the Feynman rules for the scalar quantities
$G_{0}$, $\Psi$ and $\Psi^{*}$ are modified as follows:
(i) $G_{0}(i,j)\!\rightarrow\!\underline{G}_{0}(i,j)$,
$\Psi({\bm r})\!\rightarrow\!{\bm \Psi}({\bm r})$,
and $\Psi^{*}({\bm r})\!\rightarrow\!{\bm\Psi}^{\dagger}({\bm r})$;
(ii) $\underline{A}^{\nu}$ for the vertex;
(iii) take ${\rm Tr}$ for every closed particle line
and perform summations over $\nu$.

Now, we can directly extend the consideration of Popov\cite{Popov87}
on the general structure of the perturbation expansion
to the spin-$1$ system.
We thereby conclude that eq.\ (\ref{sub-cond})
is equivalent to the extremal property:
\begin{equation}
\frac{\delta\Omega}{\delta {\bm \Psi}^{*}({\bm r})}={\bm 0}\, .
\label{dOmega-dPsi}
\end{equation}
It is also straightforward to prove the following relation:
\begin{subequations}
\label{d^2Omega-d^2Psi}
\begin{equation}
\frac{\delta^{2}\Omega}{\delta \Psi_{\sigma'}({\bm r}')
\delta\Psi_{\sigma}^{*}({\bm r})}=
\Sigma_{\sigma\sigma'}({\bm r},{\bm r}';\omega_{n}=0)
-
\mu\delta_{\sigma\sigma'}\delta({\bm r}-{\bm r}'),
\end{equation}
\begin{equation}
\frac{\delta^{2}\Omega}{\delta \Psi_{\sigma'}^{*}({\bm r}')
\delta\Psi_{\sigma}^{*}({\bm r})}=
\Delta_{\sigma\sigma'}({\bm r},{\bm r}';\omega_{n}=0).
\end{equation}
\end{subequations}
We finally consider the uniform condensate, write
the condensate wave function as eq.\ (\ref{Psi-uniform}),
and expand
$\underline{\Sigma}$ and $\underline{\Delta}$ 
as eq.\ (\ref{rho-uniform}).
It is then evident that the thermodynamic potential depends
on $\bm\Psi$ and $\bm\Psi^{\dagger}$ via the product
$n_{0}\equiv \bm\Psi^{\dagger}\bm\Psi$.
Hence we can write eqs.\ (\ref{dOmega-dPsi})
and (\ref{d^2Omega-d^2Psi}) for the uniform case as
\begin{equation}
\frac{\partial\Omega}{\partial \Psi_{\sigma}^{*}}
=\Psi_{\sigma}\frac{\partial\Omega}{\partial n_{0}} ,
\label{dOmega-dPsi-2}
\end{equation}
and 
\begin{subequations}
\label{d^2Omega-d^2Psi-2}
\begin{equation}
\frac{\partial^{2}\Omega}{\partial \Psi_{\sigma'}
\partial\Psi_{\sigma}^{*}}
=\delta_{\sigma\sigma'}\frac{\partial\Omega}{\partial n_{0}}
+\Psi_{\sigma'}^{*}\Psi_{\sigma}
\frac{\partial^{2}\Omega}{\partial n_{0}^{2}},
\end{equation}
\begin{equation}
\frac{\partial^{2}\Omega}{\partial \Psi_{\sigma'}^{*}
\partial\Psi_{\sigma}^{*}}=\Psi_{\sigma'}\Psi_{\sigma}
\frac{\partial^{2}\Omega}{\partial n_{0}^{2}},
\end{equation}
\end{subequations}
respectively.
Using eqs.\ (\ref{dOmega-dPsi})-(\ref{d^2Omega-d^2Psi-2}), we arrive at
the generalized Hugenholtz-Pines theorem:
\begin{eqnarray}
&&\hspace{-10mm}
\sum_{\sigma'}\bigl[\Sigma_{\sigma\sigma'}
({\bm k}\!=\!{\bm 0},\omega_{n}\!=\!0)
\Psi_{\sigma'}-\Delta_{\sigma\sigma'}({\bm k}\!=\!{\bm 0},\omega_{n}\!=\!0)
\Psi_{\sigma'}^{*}\bigr]
\nonumber \\
&&\hspace{-10mm}
=\mu\Psi_{\sigma}.
\label{HP-multi}
\end{eqnarray}
It tells us that there is at least a single gapless one-particle excitation
in the spin-$1$ BEC; see the discussion below eq.\ (\ref{eq:BdG-k})
on this point. It is also straightforward to extend the consideration
to any multi-component system, which leads to the same conclusion.


\end{document}